\documentclass[a4paper, 11pt]{article}
\usepackage[utf8x]{inputenc}
\usepackage[french, english]{babel}
\usepackage{amsfonts, amssymb, amsmath, amsthm, stmaryrd, ytableau,
  genyoungtabtikz, cancel, dutchcal, mathrsfs}
\usepackage{pifont, indentfirst, dsfont, relsize}
\usepackage{geometry, graphicx, tabularx}
\usepackage{fancyhdr}
\usepackage{color}
\definecolor{linkcolor}{rgb}{0,0,0.6}
\usepackage[	pdftex,colorlinks=true,	
			pdfstartview=FitV,
			linkcolor= linkcolor,
			citecolor= linkcolor,
			urlcolor= linkcolor,
			hyperindex=true,
			hyperfigures=false]
			{hyperref}

\usepackage{lastpage}
\date{}
\newcommand{\so}{\mathfrak{so}(2, D-1)}
\newcommand{\lo}{\mathfrak{so}(1, D-1)}

\newcommand{\N}{\mathbb{N}}

\newcommand{\viel}{\bar e}
\newcommand{\bnabla}{\bar \nabla}

\usepackage{color}
\definecolor{rougef}{rgb}{0.56,0,0}
\definecolor{vertf}{rgb}{0,0.5,0}
\definecolor{bleuf}{rgb}{0,0,0.8}

\begin{document}

\begin{center}
  \textbf{\Large A note about a pure spin-connection formulation of
  \\ General Relativity and spin-2 duality in (A)dS
  } \\
\end{center}

\vspace*{.8cm}

\begin{center}
Thomas Basile$^{a,b,}$\footnote{E-mail address: \href{mailto:thomas.basile@umons.ac.be}{\tt{thomas.basile@umons.ac.be}}}, 
{Xavier Bekaert$^{a,}$\footnote{E-mail address: \href{mailto:xavier.bekaert@lmpt.univ-tours.fr}{\tt{xavier.bekaert@lmpt.univ-tours.fr}}}
and 
Nicolas Boulanger$^{b,}$\footnote{Research Associate of the Fund for 
Scientific Research$\,$-FNRS (Belgium); \href{mailto:nicolas.boulanger@umons.ac.be}{\tt{nicolas.boulanger@umons.ac.be}}}
 }
\end{center}

\vspace*{.5cm}

\begin{footnotesize} 
\begin{center}
$^a$
Laboratoire de Math\'ematiques et Physique Th\'eorique\\
Unit\'e Mixte de Recherche $7350$ du CNRS\\
F\'ed\'eration de Recherche $2964$ Denis Poisson\\
Universit\'e Fran\c{c}ois Rabelais, Parc de Grandmont\\
37200 Tours, France \\
\vspace{2mm}{\tt \footnotesize }
$^{b}$
Groupe de M\'ecanique et Gravitation\\
Service de Physique Th\'eorique et Math\'ematique\\
Universit\'e de Mons -- UMONS\\
20 Place du Parc\\
7000 Mons, Belgique\\
\vspace*{.3cm}
\end{center}
\end{footnotesize}
\vspace*{.4cm}

\begin{abstract}
We investigate the problem of finding a pure spin-connection formulation of General Relativity 
with non-vanishing cosmological constant. 
We first revisit the problem at the linearised level 
and find that the pure spin-connection, quadratic 
Lagrangian, takes a form reminiscent to Weyl gravity, 
given by the square of a Weyl-like tensor.
Upon Hodge dualisation, we show that the dual gauge field
in (A)dS$_D$ transforms under $GL(D)$ in the same representation 
as a massive graviton in the flat spacetime of the same dimension.
We give a detailed proof that the physical degrees 
of freedom indeed correspond to a massless graviton 
propagating around the (anti-) de Sitter background
and finally speculate about a possible nonlinear pure-connection 
theory dual to General Relativity with cosmological constant.
\end{abstract}

\thispagestyle{empty}
\newpage

\setcounter{page}{1}

\section{Introduction}\label{sec:Intro}

The history of pure-connection formulation of General 
Relativity (GR) is old, starting with works of Eddington 
and Schr{\"o}dinger \cite{Eddington:1920, Schrodinger:1950} --- see \textit{e.g.} Section 2 of \cite{Herfray:2015rja}, for a state of the art of pure-connection formulations of gravity. 
On the one hand, in the case of Eddington--Schr{\"o}dinger's 
proposal, 
the connection is torsionless but admits nonmetricity off-shell. 
The Lagrangian density is proportional to the square-root of 
the determinant of the Ricci tensor. 
For dimensional reason, the Lagrangian density 
must be divided by the cosmological constant 
(we will always work in the geometric unit system: $c=G=1$) 
in four spacetime dimension. 
Upon identifying
the metric with the Ricci tensor of the curvature for 
the connection divided by the cosmological constant, 
the field equations state that the 
connection is metric-compatible. 
Clearly, this action and identification are only well-defined for a nonvanishing cosmological constant $\Lambda\neq 0$.
The on-shell value of Eddington--Schr{\"o}dinger's action is
equal to the product of $\Lambda$ with the spacetime volume, 
as is the case for the Einstein--Hilbert action 
with cosmological constant. 
The Palatini action plays the r\^ole of a first-order parent 
action for both of these second-order actions. 
In fact, when $\Lambda\neq 0$ the metric is an auxiliary field and
the Eddington--Schr{\"o}dinger action can be obtained 
by solving the field equation for 
the metric in terms of the connection inside the Palatini action.
\\

On the other hand, gravity 
can also be geometrically formulated 
\`a la Cartan, with vielbein (\textit{i.e.} an orthonormal frame) 
and spin-connection (\textit{i.e.} a metric-compatible connection admitting
torsion off-shell) as dynamical variables. 
For GR with nonvanishing cosmological constant, 
the Lagrangian density in the Cartan formulation 
admits\footnote{In the flat case, see \cite{Chamseddine:1976bf}.} 
a Yang--Mills-like form, \textit{i.e.} quadratic in the curvature, via
the MacDowell--Mansouri action \cite{MacDowell:1977jt} which is also a parent action for gravity.
One can in principle integrate out the vielbein
so as to reach a final action that would be 
of pure spin-connection type.\footnote{In dimension 3, 
see 
\cite{Peldan:1991mh,Julia:2005}.}
In dimensions $D>3$, this way of proceeding is involved
and, 
for technical reasons, does not lead to a closed expression for 
the fully nonlinear Lagrangian density for the spin-connection.
However, it can be obtained perturbatively 
around any solution, for instance: (anti-) de Sitter background, as was done 
till cubic interaction level in the interesting paper 
\cite{Zinoviev:2005qp}. 
\\

In this note we do not claim to reach the corresponding 
explicit form of the  fully nonlinear,
pure spin-connection formulation of GR with cosmological 
constant, but believe that the formulation we give at the
linearised level is simple and suggestive enough to allow 
for some progress toward the searched-for action.
With this goal in mind, we speculate about a possible parent action for GR that would be viewed on the same footing 
as the MacDowell--Mansouri action.
\\

Our Lagrangian takes a form that makes it immediately suited 
to the holographic renormalisation of the Einstein--Hilbert
action with cosmological constant discussed in \cite{Miskovic:2009bm}. This suggests that the 
resulting pure spin-connection action behaves well  
under quantisation.
One can also put the action as the integral of
a square-root featuring a self-dual two-form, 
upon adding the Pontryagin topological invariant 
to the Euclideanised action.
\\

The layout of the paper is as follows. 
In Section \ref{sec:2} we revisit the 
problem at the linearised level, integrating out the vielbein 
from the linearised MacDowell--Mansouri action. 
This leads to a quadratic Lagrangian  
for the sole spin-connection. Because we keep the 
Gauss--Bonnet term in the original MacDowell--Mansouri
action, even in $D=4$ where it is a total derivative, 
we land on a new, suggestive form for the pure spin-connection 
quadratic Lagrangian ${\cal L}(\omega)\,$. 
It turns out to be given by a density very reminiscent 
to the one describing Weyl's gravity. 
In the same section, we show that the dual graviton in 
the (anti-) de Sitter geometry (A)dS$_D$ transforms in 
a representation of $GL(D)$ identical 
to the representation carried by the dual \emph{massive} 
graviton in flat spacetime of the same dimension $D\,$ 
and further explain the nature of the dual graviton in (A)dS$_D\,$.
\\

The core of the  paper, in Section \ref{sec:3}, 
consists in giving a detailed proof that the resulting 
theory propagates only the 
degrees of freedom 
for a massless graviton on (A)dS background.  
This is not obvious indeed, see e.g. 
\cite{Sezgin:1979zf} where the counting of 
degrees of freedom was studied for quadratic-curvature-like 
Lagrangians featuring both the spin-connection and vierbein. 
A recent discussion on the problem of counting degrees of freedom 
for pure spin-connection Lagrangians can be found in 
\cite{Herfray:2015rja}.
In the same section, we explain in which precise sense the field equations 
for the pure spin-connection action are equivalent to the zero-torsion 
condition, upon appropriately identifying the \emph{geometrical} vielbein, 
function of the spin-connection, around (A)dS. 
\\

From the results of the previous section, In Section \ref{sec:4} 
we speculate about a possible fully
nonlinear, alternative parent action for GR. 
The proposed action would then be viewed on the same footing as 
the MacDowell--Mansouri action, considered as a parent action
for General Relativity with cosmological constant.
The nonlinear Lagrangian we propose 
reproduces 
the quadratic Lagrangian ${\cal L}(\omega)$ to lowest order.
In the same section, we add to the proposed action the 
Pontryagin term so as to express the resulting euclideanised action
as the square root of the determinant of an antisymmetric, (anti-)selfdual 
two-form. 
We conclude the paper with a summary in Section \ref{sec:5}.

\section{Linearised gravity in the pure spin-connection form}
\label{sec:2}

\subsection{A brief review of MacDowell--Mansouri gravity}

\noindent MacDowell--Mansouri's formulation of 
gravity \cite{MacDowell:1977jt}, 
extended to any dimensions $D$ in \cite{Vasiliev:2001wa},  
is based on a quadratic action in the curvature 2-form of $\lo$ or $\so\,$, depending on the sign 
of the cosmological constant, later modified 
by Stelle and West \cite{Stelle:1979aj} who gave 
its $\so$ manifestly covariant version. 
In its original version, it reads
\begin{equation}
  S_{MM}[e,\omega] = \tfrac{1}{2\lambda^2}\,\int_{\mathscr{M}_{D}} \epsilon_{abcd\,k_1 \dots k_{D-4}}
  {\cal R}^{ab}(e,\omega) \wedge {\cal R}^{cd}(e,\omega) 
  \wedge e^{k_1} \dots \wedge e^{k_{D-4}}\ ,
\label{McDM}
\end{equation}
The Lorentz-valued components of the (A)dS$_{d+1}$
curvature 2-form reads 
\begin{equation}
{\cal R}^{ab} = d \omega^{ab} +
\omega^a{}_c \wedge \omega^{cb} + \sigma \,\lambda^2 \,e^a \wedge e^b \ , 
\label{AdScurv}
\end{equation}
with $\omega^{ab}$ and $e^a$ the spin connection 
and vielbein 1-forms, respectively. The inverse
of the (A)dS radius squared, $\lambda^2$,
 is related to the cosmological constant by
$\lambda^2 = -\sigma\,\frac{2\Lambda}{(D-1)(D-2)}
$.
The parameter $\sigma = +1$ corresponds to AdS$_D$ whereas 
$\sigma = -1$ corresponds to dS$_D\,$.
In the following, for the sake of 
definiteness, we will take $\sigma=+1$ with the 
understanding that the 
results apply to dS$_D$ upon changing 
the sign in front of $\lambda^2\,$. 
The translation-valued components of the 
(A)dS$_D$ curvature 2-form coincides with the
torsion 
\begin{equation}
T^a = de^a + \omega^{a}{}_b \wedge e^b\ .
\end{equation}
The above action \eqref{McDM} is invariant under diffeomorphisms 
and local Lorentz transformations, under which 
both the vielbein and spin-connection transform 
in the usual way. \\

In $D=4\,$, the above Lagrangian density contains a 
total derivative, related to the Gauss--Bonnet invariant
\begin{equation}
I_4[\omega] = \tfrac{1}{2}\,\int_{\mathscr{M}_{4}}
\epsilon_{abcd}\;
  {R}^{ab}(\omega) \wedge {R}^{cd}(\omega)\ , 
\end{equation}
a topological invariant proportional to the Euler 
characteristic $\chi({\mathscr{M}_{4}})\,$. Explicitly, one has  
$I_4[\omega] = 16\pi^2 \chi({\mathscr{M}_{4}})\,$, where 
${R}^{ab}(\omega) = d \omega^{ab} +
\omega^a{}_c \wedge \omega^{cb} $ is the Lorentz curvature
2-form. Therefore, in $D=4\,$ one can drop it and 
obtain the Cartan--Weyl action with cosmological 
constant 
\begin{equation}
S_{CW}[e,\omega] = \int_{\mathscr{M}_{4}} 
  \left({R}^{ab}(\omega)+\tfrac{1}{2}\lambda^2  e^{a} e^{b}\right) e^{c} e^{d}\,\epsilon_{abcd}\ ,
\end{equation}
where from now on we omit the wedge product 
between differential forms. 
The equations of motion derived from $S_{CW}[e,\omega]$ are:
\begin{equation}
\epsilon_{abcd} \, T^{c}\, e^{d}\ \approx 0 \ ,
\label{eom_torsion}
\end{equation}
and
\begin{equation} 
\epsilon_{abcd} \,e^b \,{{\cal R}^{cd}}\ \approx 0 \ ,
\label{eom_curvature}
\end{equation}
where the weak equality symbol $\approx$ will stand for any equality only valid when the equations of motion hold.
The first set of field equations puts the torsion to 
zero, and provided that the components of the vierbein are invertible, 
allows one to express the spin connection in terms 
of the vierbein. The remaining field equations 
then provide vacuum Einstein's equations with cosmological 
constant $\Lambda$:
\begin{equation}
  R_{\mu \nu} - \frac{1}2 g_{\mu \nu} R + \Lambda g_{\mu \nu} \approx 0\ ,
  \label{einstein}
\end{equation}

As clearly recalled in \cite{Wise:2006sm
}, gravity in this 
formulation is made more similar to a gauge theory.\footnote{Nevertheless, 
this tantalising similarity calls for several important caveats (see e.g. 
Appendix A.3 in \cite{Bekaert:2005vh}).} 
Indeed, the action 
amounts to a Yang--Mills-like action, 
in the sense that it is quadratic in the Lorentz-valued part of the
(A)dS$_D$ curvature 2-form. 
However, a crucial difference is that it is not of the Yang--Mills
type $\int \text{tr}(F \wedge * F)$
but rather of the form $\int \text{tr}(F \wedge F)\,$. 
Another crucial difference with 
respect to the usual Yang--Mills gauge theories is that the 
underlying geometry is not described in terms of a  principal $G$-bundle, 
where $G$ is the isometry group of the maximally-symmetric spacetime,
but is a Cartan geometry (see \cite{Hehl:1994ue,Sharpe:1997} for more 
details) since, among other things, the vielbein is nondegenerate.\\

In dimension $D>4\,$, 
the term in the Lagrangian 
that is quadratic in the Lorentz curvature 2-form $R^{ab}$ is 
no longer a total derivative, but nevertheless 
the linearised field equations in $D>4$
still remain equivalent to the Fierz--Pauli equations that 
propagate a massless spin-2 field around AdS$_D$ --- see e.g. Section 2.2 in \cite{Bekaert:2005vh}.\\

At this stage, one could try and integrate out
the vielbein from the second set of field equations
\eqref{eom_curvature} and obtain the pure 
spin-connection action $S_{MM}[e(\omega),\omega]\,$. 
This
action is perfectly well-defined, although obtaining 
its explicit form remains an open problem, 
since solving \eqref{eom_curvature} in terms of the vielbein
turns out to be technically very involved.
We prefer, instead, to integrate out 
the vielbein from the linearisation of the action 
\eqref{McDM}, and then seek for nonlinear 
extensions of the resulting, pure spin-connection 
quadratic action.

\subsection{Pure spin-connection action for linearised gravity around AdS$_D$}

\noindent Linearising  MacDowell--Mansouri's action \eqref{McDM}  
around AdS$_D$ gives:
\begin{equation}
  S_\lambda[h, \omega] = \tfrac{1}{2\lambda^2}\, \int_{AdS_D}
  \epsilon_{abcd\,k_1 \dots k_{D-4}} \left(\bnabla \omega^{ab} 
  + 2\,\lambda^2\, \viel^{a} \, h^{b} \right) 
  \left(\bnabla \omega^{cd} + 2\, \lambda^2\, 
  \viel^{c}\, h^{d} \right) \viel^{k_1} \dots \viel^{k_{D-4}}
  \label{lin_mm_action}
\end{equation}
with $\bnabla = d + \bar \omega$ the 
Lorentz covariant derivative of the AdS$_D$ background, 
$\viel^a$ and $\bar \omega^{ab}$ being
respectively the background vielbein and spin connection, 
obeying $\overline{\cal R}^{ab} = d\bar \omega^{ab} 
+ \bar \omega^a{}_c \, \bar \omega^{cb} 
+ \lambda^2 \, \bar e^a \, \bar e^b = 0$ and 
$\overline{T}^a = d\bar e^a + \bar \omega^a{}_b \, \bar e^b = 0\,$, 
and $h^a$ and $\omega^{ab}$
their respective fluctuations; \textit{i.e.} $e=\bar e+h$ while, 
with some abuse of notation, we use the same symbol $\omega$ 
for the full spin connection and its fluctuation around the background. 
The linearised action is invariant under the following 
gauge transformations:
\begin{equation}
  \begin{aligned}
    \delta_\epsilon h^a = \bnabla \epsilon^a - \viel_b \, \epsilon^{ab} \ ,
    & \hspace{1.5cm} & \delta_\epsilon \omega^{ab} = \bnabla
    \epsilon^{ab} + 2\lambda^2 \, 
    \viel^{[a} \epsilon^{b]}\ ,
    \label{gaugetransfo}
  \end{aligned}
\end{equation}
where curved (square) brackets surrounding indices denote
(anti)symmetrisation with strength one. \\

\noindent The action  \eqref{lin_mm_action} leads to 
the equations of motion:
\begin{equation}
  \frac{\delta S_\lambda}{\delta h^d} \approx 0 \quad \Leftrightarrow\quad 
  \epsilon_{abcd\,k_1 \dots k_{D-4}} \left(\bnabla \omega^{ab}\, \viel^c +
  2\lambda^2 \, \viel^a \, \viel^b \, h^c \right) \viel^{k_1} \dots 
  \viel^{k_{D-4}} \approx 0 \ , 
\end{equation}
whose solution reads:
\begin{equation}
  h_{\mu}{}^a \approx -\frac{1}{(D-2)\lambda^2} \,
  \left( R_{\mu}{}^a - \frac{1}{2(D-1)}\, R \,\viel_{\mu}{}^a \right)\ ,
  \label{pseudoSchouten}
\end{equation}
where we defined 
\begin{equation}
R_{\mu \nu}{}^{ab} \,=\, 2\,\bnabla_{[\mu}\omega_{\nu]}{}^{ab} \ , 
\qquad R_{\mu}{}^a \,=\, \viel_b^\nu \,R_{\mu\nu}{}^{ab} \ , 
\qquad 
{\rm {and}} \qquad R \,=\, \viel_a^\mu\, R_{\mu}{}^a \ , 
\end{equation}
that we will call hereafter respectively Riemann-like tensor, 
Ricci-like tensor and Ricci-like scalar,  
because of their formal ressemblance with the eponymous
tensors and scalar. 
Using the above expression for $h_{\mu}{}^a\,$, 
the resulting child action can be written as:
\begin{eqnarray}
  S_\lambda[\omega] & = & -\tfrac{(D-4)!}{2\lambda^2}\, \int_{AdS_D}
  \viel \, d^Dx \ \left( R_{cd}{}^{ab} R_{ab}{}^{cd} - \frac{4}{D-2}
  R_{b}{}^a R_a{}^{b} + \frac{2}{(D-1)(D-2)} R^2 \right) \ ,
  \nonumber \\ 
& = &  -\tfrac{(D-4)!}{2\lambda^2}\, \int_{AdS_D} \viel \, d^Dx \ 
  C_{cd}{}^{ab}(\omega) C_{ab}{}^{cd}(\omega) \ ,
  \label{dual_lin_action}
  \end{eqnarray}
featuring the Weyl-like tensor 
\begin{equation}
C_{cd}{}^{ab}(\omega) := R_{cd}{}^{ab} - \frac{2}{D-2}\,
(\delta^a{}_{[c}R_{d]}{}^b-\delta^b{}_{[c}R_{d]}{}^a) +
\frac{2}{(D-1)(D-2)}\, R \ \delta^a{}_{[c}\delta^b{}_{d]} \ . 
\label{LineWeyl}
\end{equation}
Let us stress that, even though \eqref{LineWeyl} is 
formally the same expression as the usual Weyl tensor, 
it is instead defined in terms of the Riemann-like tensor, 
Ricci-like tensor and Ricci-like scalar. 
The Weyl-like tensor is traceless but not
symmetric under the exchange of the 
two pairs of upper and lower antisymmetric indices. 
In fact, by noticing that the solution \eqref{pseudoSchouten} is 
nothing but $h_{\mu}{}^a = - \tfrac{1}{\lambda^2}\,P_{\mu}{}^a$
for the Schouten-like tensor $P_{\mu}{}^a:=\frac{1}{(D-2)} \,
  \left( R_{\mu}{}^a - \frac{1}{2(D-1)}\,
  {\bar e}_{\mu}{}^a\,R \,\right)\,$, the 
expression \eqref{LineWeyl} can be expressed more compactly in 
terms of the following 
2-form:
\begin{equation}\label{Cab}
C^{ab}(\omega) = R^{ab}(\omega) + 2\lambda^2\, \viel^{[a}\, h^{b]}(\omega)\ .
\end{equation}
Comparing this expression with the linearisation of 
the AdS$_{D}$ curvature \eqref{AdScurv}, 
we readily see that, on the solution 
\eqref{pseudoSchouten} for the vielbein in terms of the 
spin-connection, 
the actions \eqref{lin_mm_action} and 
\eqref{dual_lin_action} are indeed the same. \\

\noindent The field equations derived 
from \eqref{dual_lin_action} read:
\begin{eqnarray}
  \frac{\delta S_\lambda[\omega]}{\delta \omega_{\mu}{}^{ab}} \approx 0
  & \Leftrightarrow & \bnabla_{\nu} R_{ab}{}^{\mu \nu} +
  \frac{2}{D-2} \,\left( \bnabla_{[a} R_{b]}{}^{\mu} - \viel^{\mu}{}_{[a|}
    \bnabla_{\nu} R_{|b]}{}^{\nu} \right) + \frac{2}{(D-1)(D-2)}\,
  \viel_{[a}{}^{\mu}\bnabla_{b]} R \approx 0 
  \nonumber \\ & \Leftrightarrow &
  \widetilde{C}_{ab\vert}{}^{\mu} := 
  \tfrac{1}{D-3}\,\bnabla_{\nu} C_{ab}{}^{\mu \nu} 
  \approx 0\ .
  \label{eomCotton}
\end{eqnarray}
The left-hand side of the field equations features the 
Cotton-like tensor to which we will return in the next section.
We will also show in which sense the above field equations can 
be viewed as the zero$\,$-torsion condition for the spin-connection. 
Due to the gauge symmetries
of the action under \eqref{gaugetransfo}, the left-hand 
side of the above field equations obey the following 
Noether identities:
\begin{equation}
\bnabla_{\mu}\bnabla_{\nu} C_{ab}{}^{\mu \nu} \equiv 0 
\,, \qquad 
\bnabla_{\nu} C_{ab}{}^{a \nu} \equiv 0 \ .
\label{Noether}
\end{equation}
They 
are simple consequences  of the 
tracelessness of the Weyl-like tensor. 

\subsection{Dual graviton in (A)dS$_D$}

\noindent 
Hodge duality for linearised gravity around AdS background was discussed, 
for example, in \cite{Julia:2005ze} for the Hamiltonian formulation, 
and \cite{Matveev:2004ac} for the Lagrangian formulation.
In this subsection, and more fully in the Section \ref{sec:3},
we clarify the nature of the degrees of freedom and the 
$GL(D)$ symmetry of the dual graviton in AdS$_D\,$, from the 
Lagrangian and manifestly Lorentz-covariant point of view.\\

As we have just commented above, the child action $S_\lambda[\omega]$ 
derived in \eqref{dual_lin_action}  
from the parent action \eqref{lin_mm_action} inherits the gauge symmetry  
\begin{equation}
\delta_\epsilon \omega^{ab} = \bnabla \epsilon^{ab} + 2\lambda^2 \, \viel^{[a} \epsilon^{b]}\ ,
\end{equation}
for the remaining gauge field $\omega\,$.
Again, as long as the cosmological constant is nonvanishing, 
the second term on the right-hand side above is nonzero, 
which implies that one can 
gauge fix the trace of $\omega_\mu{}^{\alpha\beta}$ to zero. 
As a consequence, in the gauge where $\omega_\mu{}^{\mu\beta}\equiv 0\,$,  
dualising the pair of upper antisymmetric indices of
$\omega_\mu{}^{\alpha\beta}$ in (A)dS$_D$ gives a dual potential
\begin{equation}
\tfrac{1}{2}\,\varepsilon_{\alpha\beta\alpha_1\ldots \alpha_{D-2}} 
\,\omega_\mu{}^{\alpha\beta}
\, =: \, \widetilde{\omega}_{\mu \vert \alpha_1\ldots \alpha_{D-2}}
\;=\; \widetilde{\omega}_{\mu \vert [\alpha_1\ldots \alpha_{D-2}]}
\label{dual1}
\end{equation}
that transforms in the hook-like, irreducible $GL(D)$ representation 
characterised by
\begin{equation}
\varepsilon^{\mu \nu \alpha_1\ldots \alpha_{D-2}}\,\widetilde{\omega}_{\mu 
\vert \alpha_1\ldots \alpha_{D-2}}
\equiv 0\ .
\label{dual2}
\end{equation}
For example, in (A)dS$_4\,$, 
\begin{equation}
\widetilde{\omega}_{\mu \vert \alpha_1\alpha_{2}} 
\qquad \sim \qquad  
\Yboxdim{7pt} \Ylinethick{.8pt} \yng(2,1)\qquad ,
\end{equation}
whereas in (A)dS$_5\,$, 
\begin{equation}
\widetilde{\omega}_{\mu \vert \alpha_1\alpha_{2}\alpha_{3}} 
\qquad \sim \qquad  
\Yboxdim{7pt} \Ylinethick{.8pt} \yng(2,1,1)\qquad ,
\end{equation}
and so on for $D>5\,$.
We demonstrate in the next section that the theory with action 
\eqref{dual_lin_action} does describe a massless graviton around 
(A)dS$_D\,$.
Therefore we see that the $GL(D)$ representation for the 
dual graviton in (A)dS$_D\,$ differs from the representation of 
the dual graviton in $\mathbb{R}^{1,D-1}\,$ 
\cite{West:2001as,
Boulanger:2003vs}
by the presence of an extra box in the first column of its 
associated Young diagram.  
This might come as a surprise, taking into account the fact, 
explained in \cite{Brink:2000ag,Boulanger:2008up,Alkalaev:2009vm},
that a mixed-symmetry gauge field in (A)dS$_D$ 
propagates more degrees of freedom compared to the  
massless field in $\mathbb{R}^{1,D-1}\,$ associated with the same 
$GL(D)\,$ Young tableau.
In this sense, those gauge fields in (A)dS$_D$ are more akin massive fields, 
and therefore the symmetries exhibited here in 
\eqref{dual1}--\eqref{dual2} might come as a surprise, 
since the corresponding $GL(D)$ Young diagrams  
are those characterising a dual \emph{massive} graviton in flat
spacetime \cite{Curtright:1980yj}, see also 
\cite{Casini:2002jm,Zinoviev:2005zj,Gonzalez:2008ar,Morand:2012vx}. \\

The resolution of this apparent paradox precisely comes by using the 
result of the analysis of \cite{Brink:2000ag,Boulanger:2008up,Alkalaev:2009vm}: 
In the flat limit from (A)dS$_4\,$, the gauge field $\widetilde{\omega}_{\mu\vert \alpha\beta}$
decomposes as follows 
\begin{equation}
\widetilde{\omega} \quad 
\leftrightarrow \quad \Yboxdim{7pt} \Ylinethick{.8pt} \yng(2,1)\quad \underset{\lambda \rightarrow 0}{\sim}
\quad \Yboxdim{7pt} \Ylinethick{.8pt} \yng(2,1) \quad \oplus \quad 
\Yboxdim{7pt} \Ylinethick{.8pt} \yng(2) \quad ,\qquad (D=4)
\end{equation}
whereas, in the flat limit from (A)dS$_5\,$, one has 
\begin{equation}
\widetilde{\omega} \quad 
\leftrightarrow \quad \Yboxdim{7pt} \Ylinethick{.8pt} \yng(2,1,1)\quad \underset{\lambda \rightarrow 0}{\sim}
\quad \Yboxdim{7pt} \Ylinethick{.8pt} \yng(2,1,1) \quad \oplus \quad
\Yboxdim{7pt} \Ylinethick{.8pt} \yng(2,1)\quad , \qquad (D=5) 
\end{equation}
and so forth for $D>5\,$. 
The first gauge field appearing on the right-hand side of the 
decompositions above is topological in flat space, so that only the second 
gauge field is propagating, which is precisely the gauge field dual 
to a massless graviton in the flat space of the corresponding
dimension \cite{Boulanger:2003vs}, thereby explaining why the 
gauge field $\widetilde{\omega}$ can propagate in (A)dS$_D$ 
the degrees of freedom of a massless graviton.

\section{Physical degrees of freedom}
\label{sec:3}

\subsection{Strategy}

In this section, we use the unfolding technology 
\cite{Vasiliev:1988sa
}\,\footnote{For a
technical review of the spin-two case, see e.g.
Section 7 of \cite{Bekaert:2005vh}
or Section 4 of \cite{Didenko:2014dwa}.}
in order to prove that the field equations \eqref{eomCotton}
indeed propagate a massless graviton on the 
AdS$_{D}$ background. 
More precisely, we refer to the work  
\cite{Iazeolla:2008ix} where the unfolding 
of linearised spin-$s$ gauge theory in AdS$_{D}\,$,  
and in particular linearised gravity for $s=2\,$, 
is explained in great details.
The 1-particle states of a physical massless spin-2 field 
around AdS$_{D}$ form an irreducible and unitary 
$\mathfrak{so}(2,D-1)$ representation that can be mapped, 
via harmonic expansion \cite{Iazeolla:2008ix}, 
to the infinite tower of Lorentz tensors 
${\cal T}$ transforming in the following $\lo$ representations
depicted by the associated Young diagrams:
\begin{equation}
  \Yboxdim{7pt} \Ylinethick{1pt} {\cal T} = \left\{ \yng(2,2)\, ,
  \ \yng(3,2)\, , \ \yng(4,2)\, , \ \yng(5,2)\, , \ \dots \right\} 
  = \left\{ \Yboxdim{7pt} \Ylinethick{1pt} \newcommand\n{s}
  \gyoung(;;_5\n,;;) \right\}_{s \in \N}\ .
\end{equation}
This infinite tower of $\lo$ tensors satisfies a first-order 
differential equation that links each of the tensors irreducibly 
by the action of the $\so$ translation
generators \cite{Iazeolla:2008ix}. 
All these $\lo$ tensors correspond 
to the on-shell Weyl tensor and all its on-shell 
nontrivial covariant derivatives. 
At one point of spacetime, the data of these tensors 
is equivalent to all the nontrivial Taylor coefficients of the 
gravitational field at this point, thereby allowing 
to reconstruct the field everywhere in an open patch. 
In other words, these Lorentz 
tensors are mapped one-to-one to the coefficients
of the metric in the normal Riemann coordinates 
expansion.\\

In order to prove that our pure spin-connection 
formulation of linearised gravity with cosmological 
constant correctly 
describes the propagating massless spin-2 field, 
we thus have to show that the only gauge-invariant 
tensors that are not constrained by the EOM 
\eqref{eomCotton}
correspond to the various projections of the covariant derivatives 
of the Riemann-like tensor $R_{ab | cd}$ 
on the symmetries of the $\lo$ tensors displayed 
in ${\cal T}\,$. 
In order to do so, we  start by comparing the
Lorentz projections of the Riemann-like tensor, together 
with all its AdS$_{D}$ covariant derivatives, with 
the corresponding projections of the derivatives of the 
EOM in order to eliminate those components that vanish by virtue 
of \eqref{eomCotton}.\\

The outcome of this analysis will be that, indeed, 
the only Lorentz-irreducible projections of the 
successive 
covariant derivatives of the Riemann-like tensor
that are (i) gauge-invariant and (ii) nonvanishing 
on the solutions of \eqref{eomCotton},
are in one-to-one correspondence  
with the Lorentz tensors in the set ${\cal T}\,$. 
Furthermore, we will show that the first tensor in this 
set,  that we will call the \emph{primary Weyl tensor}
following \cite{Boulanger:2008up}, 
obeys the D'Alembert equation in AdS$_{D}\,$,
\begin{equation}
\Big( \Box - 2\lambda^2\,(D-1)\Big) W_{abcd} = 0\;, 
\end{equation}
which, together with the relations linking all 
the higher Lorentz tensors in ${\cal T}\,$, ensures the 
isomorphism of the $\so$ module ${\cal T}$ with the 
unitary irreducible representation of $\so$ specifying 
the massless graviton, as 
explained in details in \cite{Iazeolla:2008ix} --- see 
also \cite{Boulanger:2008up} for a review of 
linearised unfolded systems around maximally-symmetric backgrounds.

\subsection{Gauge-invariant and traceless projections of $R_{ab | cd}$}

Clearly, upon inspection of
\eqref{gaugetransfo}, there is no gauge-invariant 
quantity built out of the undifferentiated 
spin-connection. At first order in the derivatives of $\omega^{ab}\,$, 
we decompose 
$\bnabla_\mu \omega_{\nu}^{ab} = \bnabla_{(\mu} \omega_{\nu)}{}^{ab} +\bnabla_{[\mu} \omega_{\nu]}{}^{ab}\,$.  
The first piece transforms 
with the second symmetrised derivative of parameter 
$\epsilon^{ab}$ and is not invariant. We therefore 
start the analysis with the second piece 
$\bnabla_{[\mu} \omega_{\nu]}{}^{ab}\,$ which is,
up to an inessential factor of 2, the 
Riemann-like tensor, and then consider
all its symmetrised covariant derivatives. 
Recalling the expression for the Schouten-like tensor 
\begin{equation}
P_{a\vert b} = \frac{1}{(D-2)} \,
  \left( R_{a\vert b} - \frac{1}{2(D-1)}\,
  \eta_{ab}\,R \,\right)\ ,
\end{equation}
together with the decomposition given in \eqref{LineWeyl}: 
\begin{equation}
R_{cd\vert}{}^{ab} = C_{cd\vert}{}^{ab} 
+ 4 \,\delta^{[a}{}_{[c}P_{d\,]\vert}{}^{b]}\ ,  
\label{decompoRiemann}
\end{equation}
we see that considering all the symmetrised covariant derivatives 
of $R_{cd\vert ab}\,$
is equivalent to considering separately all the symmetrised 
covariant derivatives of $C_{cd\vert ab}$
and of $P_{a\vert b}\,$.
Notice that we use a vertical bar to separate 
groups of antisymmetric indices and that the background vielbein 
has been used to transform all base
indices (Greek) into fiber
ones (Latin).
Since the transformation law of the Riemann-like tensor 
under \eqref{gaugetransfo} is
\begin{equation}
\delta_\epsilon R_{ab|cd} = -2\lambda^2
  \left( \eta_{c [a} \bnabla_{b]} \epsilon_d - \eta_{d
    [a} \bnabla_{b]} \epsilon_c + \eta_{c [a}
    \epsilon_{b] d} - \eta_{d [a} \epsilon_{b] c}
  \right)\ ,
\end{equation}
we see that all its traceless projections are gauge 
invariant, hence $C_{ab|cd}$ is gauge invariant. 
As for the transformations of the Ricci-like tensor 
and Ricci-like scalar, we have  
\begin{eqnarray}
\delta_\epsilon R_{ab} &=& -\lambda^2 \eta_{ab} \bnabla_c \, 
\epsilon^c -\lambda^2 (D-2) \left( \bnabla_a \, 
\epsilon_b + \epsilon_{ab} \right)\ ,
\nonumber \\
\delta_\epsilon R &=& -2\lambda^2 (D-1) \bnabla_a \, \epsilon^a \ ,
\label{transfoRicci}
\end{eqnarray}
leading to the transformation law 
for the Schouten-like tensor:
\begin{equation}
\delta_\epsilon P_{a\vert b} = -\lambda^2 (\bnabla_a\epsilon_b + \epsilon_{ab} )\ .
\label{transfoSchouten}
\end{equation}
As none of the tensors introduced so far are irreducible under the Lorentz group, 
we proceed now to the Lorentz-irreducible decompositions 
of the undifferentiated Riemann-like tensor.  
We use $\lo$ Young diagrams to specify the various 
irreducible pieces. The list of these components is 
given in Table \ref{tab:projections}.  
\begin{table}[!ht]
\centering
\begin{tabular}{|c|c|}
\hline
Young diagrams & Corresponding tensors\\ \hline
 $\Yboxdim{7pt} \Ylinethick{.8pt} \yng(2,2)$ & 
 $W_{ab | cd} :=
  I_{ab | cd} - \tfrac{2}{D-2}\,\left(\bar g_{c[a} I_{b] | d} - \bar
  g_{d[a} I_{b] | c} \right) + \tfrac{2}{(D-1)(D-2)}\, \left( \bar
  g_{c[a} \bar g_{b]d} - \bar g_{d [a} \bar g_{b]c} \right) I\,$,  \\
  & with $I_{ab|cd} := \frac{1}6 \left(R_{ab | cd} + R_{cd | ab} 
- R_{c [a | b] d} + R_{d [a | b] c} \right)\,$, 
$I_{bd} := \frac{1}2 R_{(b | d)}$ and $I := \frac{1}2 R\,$ \\ \hline
 $\Yboxdim{7pt} \Ylinethick{.8pt} \yng(2,1,1)$ & $\hat J_{a b c | d}
  := J_{abc | d} - \frac{1}{D-2} \left(2\bar g_{c [a} J_{b] | d} 
  + \bar g_{cd} J_{a | b} \right)\,$, \\
  & with $\quad J_{abc | d} := \frac{1}2 \left( R_{[ab|c]d} - R_{d [a|bc]} \right)\quad $ 
and $\quad J_{b | d} := \frac{2}3 R_{[b | d]}\,$ \\ \hline
$\Yboxdim{7pt} \Ylinethick{.8pt} \yng(1,1,1,1)$ & $\hat K_{a b c d} = R_{[ab | cd]}$
\\ & \\ \hline
$\Yboxdim{7pt} \Ylinethick{.8pt} \yng(1,1)$ & $R_{[a|b]}$ \\ & \\ \hline
$\Yboxdim{7pt} \Ylinethick{.8pt} \yng(2)$ & $R_{(a|b)} - \tfrac{1}D \, \bar g_{ab} R$ \\ & \\ \hline
$\bullet$ & $R$ \\ \hline
\end{tabular}
\caption{Lorentz-irreducible decomposition of the 
undifferentiated Riemann-like tensor}
\label{tab:projections}
\end{table}
The first three $\lo\,$-irreducible components
of the Riemann-like tensor $R_{ab\vert cd}\,$ appearing in the table, denoted by $W\,$, 
$\hat{J}\,$ and $\hat{K}\,$, are gauge invariant. 
They are the 3 irreducible components of the Weyl-like tensor $C_{ab\vert cd}\,$.
We show in the next subsection that, from these 3 gauge-invariant 
tensors, only the first piece, $W\,$, is not zero 
on the EOM~\eqref{eomCotton}.

\subsection{Projections of the first derivative of the equations of motion} 

The left-hand side of the field equations 
\eqref{eomCotton} are gauge invariant. 
Since they start with the first derivative 
of the Riemann-like 
tensor, one could believe that all 
the gauge-invariant components of the 
undifferentiated Riemann-like 
tensor are on-shell nontrivial observables, hence should 
be part of the set $\cal T\,$ defining the 
$\so$ module carrying the physical degrees of freedom. 
However, since the 
covariant derivatives in AdS$_{D}$ do not commute to 
zero, there can be differential consequences of 
\eqref{eomCotton} that lower the derivative order by 
2 units, thereby bringing in gauge-invariant 
components of the undifferentiated Riemann-like 
tensor. We show that this is indeed the case, and that 
from the decomposition of the previous subsection, 
only the traceless tensor $W_{ab\vert cd}$ survives 
on-shell and hence enters the set $\cal T\,$ at zeroth
order in the covariant 
derivatives of the Riemann-like tensor. 
The component $W_{ab\vert cd}$ is called 
the \emph{primary Weyl tensor}, see \cite{Boulanger:2008up}. 
\\

The way to bring down by two units the number of derivatives acting on the spin-connection in the field
equation is by computing 
$\bnabla_{[d|} \bnabla_e C_{|ab]|c}{}^e \,$ and 
decomposing it under $\lo\,$. 
We find 
  \begin{equation}
     0 \approx \bnabla_{[d|} \bnabla_e C_{|ab]|c}{}^e = -\lambda^2 (D-3) \left(
        R_{[ab|d]c} + \frac{2}{D-2} \bar g_{c[a} R_{bd]} \right)\ .
        \label{relation}
\end{equation}
By virtue of the first Noether identity \eqref{Noether}, 
the above quantity is identically traceless
and implies that  both 
$\hat J_{a b c | d}$ and $\hat K_{a b c d}$ vanish 
on-shell.
On the other hand, taking the projection of 
$\bnabla_d \bnabla_e C_{ab|c}{}^e$ on the symmetries 
of the primary Weyl tensor gives 
  \begin{equation}
    \Yboxdim{3pt} \Ylinethick{.5pt} \mathbb{P}^{\yng(2,2)}(\bnabla_d
    \bnabla_e C_{ab|c}{}^e) = \bnabla_e \Yboxdim{3pt}
    \Ylinethick{.5pt} \mathbb{P}^{\yng(2,2)}(\bnabla_d
    C_{ab|c}{}^e) + \lambda^2 (D-2) W_{ab|cd}\ .
  \end{equation}
Since  the projection $\Yboxdim{3pt} \Ylinethick{.5pt} \mathbb{P}^{\yng(2,2)}(\bnabla_d
  C_{ab|c}{}^e)$ does not produce any 
  commutator of covariant derivatives acting on 
  the spin-connection, 
from the various gauge-invariant 
components of $R_{ab\vert cd}\,$, 
only $W_{ab\vert cd}$ survives on-shell, 
as announced.

\subsection{Irreducible components of $\bnabla_e R_{ab | cd}$}

No covariant derivatives 
$\bnabla_{(a_1} \ldots \bnabla_{a_k)} R$ are 
gauge invariant, see \eqref{transfoRicci}. 
Similarly, the symmetrised derivatives
of the Ricci-like tensor 
\begin{equation}
\bnabla_{(a_1} \ldots \bnabla_{a_k} R_{a_{k+1}a_{k+2})}
\end{equation}
 are not 
gauge-invariant. 
Denoting $A_{ab}:=R_{[a\vert b]}\,$, it is also simple to see that no derivatives  
\begin{equation}
\bnabla_{(a_1} \ldots \bnabla_{a_k}A_{a) b}
\end{equation}
can be completed into a gauge invariant quantity  either. 
On the other hand, though the Ricci-like tensor is not gauge 
invariant, appropriate projections of 
its covariant derivatives can be. Again, instead of using 
$R_{a\vert b}$ and $R\,$, it is better to consider the Schouten-like tensor $P_{a\vert b}$ instead. 
Introducing the following tensor
\begin{equation}
{C}_{ab\vert c} := 2 \bnabla_{[b}P_{a]c}  \ , 
\end{equation}
and recalling \eqref{transfoSchouten}, we find 
\begin{equation}
\delta_\epsilon {C}_{ab\vert c} = 
2\lambda^2 ( \bnabla_{[a}\epsilon_{b]c}  - 
\lambda^2 \eta_{c[a}\epsilon_{b]}) \ , 
\end{equation}
so that the following tensor  is gauge invariant: 
\begin{equation}
\widetilde{C}_{ab\vert c} := 
2  \,\bnabla_{[b}P_{a]c} - 2 \,\lambda^2\,\omega_{[a\vert b]c}
\equiv {C}_{ab\vert c} - 2 \lambda^2\,\omega_{[a\vert b]c}\  . 
\label{defpseudoCotton}
\end{equation}
It is however zero on-shell, 
as we anticipated with our notation in \eqref{eomCotton}.
Indeed, using 
\begin{equation}
  \bnabla_{[a} R_{bc]|de} = 2\lambda^2 \left( \eta_{d [a}
    \omega_{b|c] e} - \eta_{e[a} \omega_{b|c]d} \right)\ , 
\end{equation}
we see that the following identity is true:
\begin{equation}
\bnabla_{[a}C_{bc]\vert de} 
+ \eta_{d[a} \widetilde{C}_{bc]\vert e} 
-  \eta_{e[a} \widetilde{C}_{bc]\vert d} \;\equiv \;0\ , 
\label{BianchiforWeyl}
\end{equation}
from which, upon taking traces, we obtain 
\begin{equation}
\widetilde{C}_{ab\vert c} \equiv 
\tfrac{1}{D-3}\,\bnabla^d C_{ab\vert cd}\;\approx \; 0\ ,
\label{CottondivC}
\end{equation}
thereby justifying the identity of the tensors $\widetilde{C}$ 
appearing in \eqref{eomCotton} and \eqref{defpseudoCotton}. 
Notice that the relation \eqref{relation} can easily be derived 
from \eqref{CottondivC} and \eqref{defpseudoCotton}. \\

We can now explain in which sense the field equation \eqref{CottondivC}
can be read as a zero-torsion condition. 
If one defines, in accordance with 
\eqref{pseudoSchouten}, 
the \emph{geometric} --- or \emph{dual} --- vielbein
$\widetilde{e}^{\,a}(\omega)$ around AdS$_D$ by its components
\begin{equation}
\widetilde{e}_\mu{}^a(\omega) \;:=\; \bar{e}_\mu{}^a - \tfrac{1}{\lambda^2}\, P_{\mu\vert}{}^a + {\cal O}(\omega^2) \ , 
\label{geometricvielbein}
\end{equation}
then the field equations \eqref{CottondivC} precisely are the expression of the 
zero-torsion condition for the full spin-connection 
\begin{equation}
\nabla = d + w \ , \qquad w := \bar\omega + \omega \ ,
\end{equation}
up to first order in the fluctuation $\omega\,$:
\begin{equation}
0 \;\approx \; T_{\mu\nu}^a \;:=\; 2\,\nabla_{[\mu}\widetilde{e}{}_{\nu]}{}^a 
\;=\; \overline{T}^a_{\mu\nu} + \frac{1}{\lambda^2}\,\widetilde{C}_{\mu\nu\vert}{}^a 
+ {\cal O}(\omega^2) \ ,
\label{zerotorsion}
\end{equation}
since $\overline{T}^a_{\mu\nu}\,$, the torsion of AdS, vanishes identically.\\

We have thus shown that the gauge-invariant completion 
of the antisymmetrised derivative
$\bnabla_{[b}P_{a]\vert c}$ is actually null on-shell, whereas 
the symmetrised covariant derivatives of $\bnabla_{(a}P_{b)\vert c}$
are not gauge invariant. 
From what we have discussed above and the 
decomposition \eqref{decompoRiemann}, 
we are thus led to look at the various contributions 
of the covariant derivatives of 
$C_{ab | cd}\,$, the
traceless part of the Riemann-like tensor. 
We already know that its component $W_{ab | cd}$ is non-vanishing on-shell, 
and that in fact it is the only component of $C_{ab | cd}\,$ that does not 
vanish on-shell, as we showed that both $\hat{J}_{abc\vert d}$ and $\hat{K}_{abcd}$ 
are zero on the solutions of \eqref{eomCotton}. 
Working on-shell, to first order in the derivatives of the 
$C_{ab | cd}\,$, we have to consider the two linearly independent 
contributions $\bnabla_{[a} W_{bc]|de}$ and 
$\bnabla_{(a} W_{bc),de}\,$, where $W_{ab,cd} := W_{c(a\vert b)d}\,$, 
where we separate groups of symmetrised indices by a coma. 
It is an important consequence of the identity 
\eqref{BianchiforWeyl} that, on-shell where 
the 3 tensors $\widetilde{C}_{ab\vert c}\,$,
$\hat{J}_{abc\vert d}$ and $\hat{K}_{abcd}$
are vanishing, we have 
\begin{equation}
\bnabla_{[a}W_{bc]\vert de} \approx 0 \ .
\label{Bianchi_Weyl_on_shell}
\end{equation}
Therefore, only the component 
$\bnabla_{(a} W_{bc),de}\,$ will have to be considered. 
Acting on the left-hand side of \eqref{Bianchi_Weyl_on_shell}
with $\bnabla^a\,$ and using the algebraic symmetries 
of $W_{ab\vert cd}\,$ together with the identity
\begin{equation}
  [\bnabla_m, \bnabla_a] W_{cd|b}{}^m = \lambda^2 (D-2)\, W_{ab|cd}
  + \left( W_{ac|bd} - W_{ad|bc} \right)\ 
\end{equation}
and the on-shell equalities 
\begin{equation}
\bnabla^a W_{ab\vert cd}\approx \bnabla^a C_{cd\vert ab}\approx 0 \ ,
\label{nuldivergence}
\end{equation}
we deduce that
\begin{eqnarray}
& \Box W_{bc\vert ef} + [\bnabla^a , \bnabla_b]W_{ca\vert ef} 
+ [\bnabla^a , \bnabla_c]W_{ab\vert ef} \approx 0 \ &
\nonumber \\ 
& \Leftrightarrow & 
\nonumber \\ 
&
\Big( \Box - 2\lambda^2\,(D-1)\Big) W_{ab \vert cd} \approx 0 \ ,
&
\label{DAlembert}
\end{eqnarray}
which is the D'Alembert equation characterising a massless spin-2 field 
freely propagating on AdS$_{D}\,$, as announced in the preamble of the Section. 
The component $\bnabla_{(a} W_{bc),de}\,$ of 
the first derivative of the primary Weyl tensor is 
linearly independent from 
$\bnabla_{[a} W_{bc]\vert de}\,$. 
It is traceless on-shell, due to  
$\bnabla^a W_{ab\vert cd}\approx 0\,$, 
and its traceless part is not constrained by 
the field equations, being independent from 
$\bnabla_{[a} W_{bc]\vert de}\,$. 

\subsection{General structure}

Let us generalise what we observed in the previous 
subsections:
Suppose that, after having taken $k-1$ derivatives of
  $W_{ab|cd}\,$, 
  the only non-vanishing gauge-invariant projection remaining is
  $\Yboxdim{8pt} \Ylinethick{.8pt} \newcommand\n{k-1}
  \gyoung(;;_5\n,;;)$, i.e. all $\,\lo\,$-irreducible 
  projections containing more than 2 rows
  or more than 2 boxes in the second row are identically zero.
Then, applying $k$ symmetrised derivatives on $W_{ab|cd}\,$ will
  yield:
  \begin{equation}
    \Yboxdim{10pt} \Ylinethick{1.2pt} \newcommand\n{k}
    \newcommand\nn{k-1} \newcommand\nnn{k+1} \gyoung(_5\n) \otimes
    \yng(2,2) = \gyoung(;;_5\n,;;) \oplus \gyoung(;;_4\nn,;;,;) \ +
    \ \text{traces}
  \end{equation}
Firstly, the second Young diagram contains more than 2 rows and thus vanishes on-shell,  
  as a consequence of \eqref{Bianchi_Weyl_on_shell}. 
Secondly, the trace terms will be 
  zero, up to lower-order terms in the covariant 
  derivatives, by virtue of \eqref{nuldivergence}. 
Together with the equation \eqref{DAlembert}, 
this finishes the proof that the on-shell degrees
of freedom propagated by the pure spin-connection 
EOM \eqref{eomCotton} 
correspond to a massless spin-2 graviton 
around (A)dS$_D$ background. 

\section{Toward a nonlinear completion}
\label{sec:4}

At the free level, 
there are in general two parent actions displaying the
equivalence, \`a la Fradkin--Tseytlin \cite{Fradkin:1984ai}, 
of two dual action principles; see \cite{Boulanger:2003vs}
in the context of linearised gravity around Minkowski 
background.
In this section, we discuss a possible 
nonlinear completion of the pure 
spin-connection action \eqref{dual_lin_action}. 
One might speculate that this functional could play 
the r\^ole of an alternative parent action, 
to be considered alongside the first-order MacDowell--Mansouri action. 

\subsection{Discussion about a nonlinear completion}

At this stage, it is tempting to consider the following
first-order action
\begin{equation}
  S[\omega, {e}] = -\frac{1}{2\lambda^2} \int_{\mathscr{M}_D}
  \epsilon_{abcd\,k_1 \dots k_{D-4}} C^{ab} C^{cd} {e}^{k_1} \dots {e}^{k_{D-4}}
  \label{nonlinearCC}\ ,
\end{equation}
where this time, 
$  C^{ab}({e},\omega) = R^{ab} - 2 \, {e}^{[a} P^{b]}\,$ 
is the full Weyl 2-form with 
$P^{a} =
  \frac{1}{D-2} \left(R^a - e^a \frac{R}{2(D-1)} \right)$
 the complete Schouten 1-form.
The Ricci 1-form is defined in terms of 
the inverse vielbein $E^\mu{}_a\,$, viz. 
$R_\mu{}^a = E^\nu{}_b R_{\mu \nu}{}^{ab}\,$, 
and the same for the Ricci
scalar $R = E^\mu{}_a R_\mu{}^a\,$.
We warn the reader that, although we use the same 
symbols as in the previous section, the above quantities 
are the full, non-linear ones. 
The general context of the discussion in the present 
section should prevent any confusion.\\

Despite its formal resemblance of \eqref{nonlinearCC} 
with the frame formulation \cite{Kaku:1977pa} of Weyl's gravity, 
the latter two action principles are inequivalent.
In fact, the action principle in \cite{Kaku:1977pa} is actually of second 
order in the vierbein as it assumes that the spin connection $\omega$ is 
expressed in terms of the vierbein $e\,$ via the condition that the torsion vanishes.
To stress again the difference between \eqref{nonlinearCC}
and Weyl's gravity, notice that, 
even in $D=4\,$, the nonlinear parent action \eqref{nonlin_eom} does 
\emph{not} enjoy the Weyl symmetry
\begin{equation}
\delta_{\sigma} \omega_\mu{}^{ab} = 2 e_\mu{}^{[a} e^{b]\nu} \,\partial_{\nu}\sigma \ , \qquad
\delta_{\sigma} e^a = \sigma\,e^a \ , \qquad
\end{equation}
since
\begin{equation}
\delta_\sigma C_{\mu \nu}{}^{ab} =  2\, T_{\mu \nu}{}^{[a} e^{b] \rho} \partial_\rho \sigma \ .
\end{equation}

\noindent The
equation of motion $\tfrac{\delta S}{\delta e^m_\lambda}\approx 0$ for the
action \eqref{nonlinearCC} are (for $D \geqslant 4\,$), 
\begin{equation}
  \epsilon^{\mu \nu \rho \sigma \, \tau_1 \dots \tau_{D-4}} \epsilon_{abcd\, k_1 \dots k_{D-4}} 
  \left( \delta_\mu^\lambda \delta_m^a \, R_\nu{}^b \, C_{\rho \sigma}{}^{cd} \, e^{k_1}_{\tau_1}
  + (D-4) \delta_{\tau_1}^\lambda \delta_m^{k_1} \, C_{\mu \nu}{}^{ab} \, C_{\rho \sigma}{}^{cd} \right) 
  e_{\tau_2}^{k_2} \dots e_{\tau_{D-4}}^{k_{D-4}} \approx 0 \, .
  \label{nonlin_eom}
\end{equation}
whereas $\tfrac{\delta S}{\delta \omega_\mu{}^{ab}}\approx 0$ yields:
\begin{equation}
  \epsilon^{\mu\nu\rho\sigma \, \tau_1 \dots \tau_{D-4}} \epsilon_{abcd \, k_1 \dots k_{D-4}} 
  \left( \nabla_\nu C_{\rho\sigma}{}^{cd}\,e_{\tau_1}^{k_1} + 2\, (D-4)\, C_{\rho\sigma}{}^{cd}\, T^{k_1}_{\nu\tau_1}\right)
  e_{\tau_2}^{k_2} \dots e_{\tau_{D-4}}^{k_{D-4}}  \approx 0 \ ,  
  \label{deltaomega_ab}
\end{equation}
Notice that the above equation does not imply that the torsion 
$T^a := de^a + \omega^a{}_b\,e^b$ vanishes on-shell. 
In this dual picture where the postulated parent action is 
\eqref{nonlinearCC}, there is no reason to expect that the fields
that we denoted by $e_\mu^a$ should be identified with the geometric  
vielbein.\\

We expect that, as in the linearised case studied in the previous section,
the field equations for the fields $e_\mu^a$ enable
one to identify the geometric vielbein with a function of the spin-connection
that, around AdS$_D\,$, starting with the Schouten tensor, see 
\eqref{geometricvielbein}. 
And similarly to the linearised case, we expect that the field equation
\eqref{deltaomega_ab} for the spin-connection should amount to the vanishing-torsion
condition for the geometric vielbein, as we have shown in  
\eqref{zerotorsion}.
The torsion $T^a$ for $e^a$ will appear, as for example in the nonlinear 
generalisation of the identity \eqref{BianchiforWeyl}:%
\begin{equation}
\nabla C^{ab} + 2 T^{[a}\,P^{b]} + 2 e^{[a}\,\widetilde{C}^{b]} \equiv 0 \ ,
\quad {\rm{where}}\quad \widetilde{C}^a := - \nabla P^{a}\ .
\end{equation}

In $D=4$ dimensions and starting from the action 
\eqref{nonlinearCC}, one readily sees that 
the second term in the field equations for the vielbein is absent 
so that the latter reduce to 
\begin{equation}
\epsilon^{\mu \nu \rho \sigma} \epsilon_{abcd} P_\nu{}^b
  C_{\rho \sigma}{}^{cd} \approx 0 \qquad 
  (D=4) \ .
  \label{vielbeinEOMfull}
\end{equation}

On top of the trivial solution $C_{\mu \nu}{}^{ab} = 0$ 
which covers the conformally flat spaces, the above 
equation also admits an obvious class of solutions, 
namely those of the type 
\begin{equation}
P_\mu{}^a = \frac1{2(D-1)f(x)}  \,\, e_\mu{}^a \quad 
\Leftrightarrow \quad 
R_\mu{}^a = \frac1{f(x)} \, e_\mu{}^a \ , 
\label{Einsteinlike}
\end{equation}
for any (smooth) nowhere vanishing 
function $f(x)\,$. This class of solutions
includes Einstein-like\footnote{We use this terminology, 
in accordance with the linearised analysis made above, 
to stress that the torsion $T^a$ is not necessarily 
zero.} spaces for constant functions 
$f(x) = k\neq 0$, and, \textit{a fortiori},
\rm \ maximally symmetric spaces.

%
Notice that these two classes of solutions are not disjoint. 
More precisely, in the case where the torsion $T^a$ is zero, 
a spacetime is both conformally flat and Einstein if 
and only if it is of constant curvature (since the Ricci scalar is the only 
nonvanishing Lorentz-irreducible component of the Riemann curvature, 
in which case the Ricci scalar must be constant by virtue of 
the Bianchi identity).
In this respect, maximally-symmetric spacetimes such as (A)dS are the 
simplest spaces in the intersection between these two classes.\\

Disregarding the conformally flat solutions of the EOM
\eqref{vielbeinEOMfull} which do not fall into the class
$R_\mu{}^a = \frac1{f(x)} \, e_\mu{}^a\,$, we can solve 
perturbatively, around AdS$_{D}\,$, the  equation 
\eqref{vielbeinEOMfull}  for the vierbein in terms
of the spin-connection. Of course, one cannot have the 
solution in closed form, but only order by order.
At the lowest order in expansion and for the function 
$f \,=\, - \frac1{(D-1)\,\lambda^2}\, = \,\frac1{2\Lambda/(D-2)}\,$, 
\textit{i.e.}
the constant corresponding to AdS$_D\,$,  
the action \eqref{nonlinearCC} reproduces 
\eqref{dual_lin_action}.
It seems that the cubic part of the pure spin-connection 
action thereby obtained by substituting the vierbein in 
terms of the spin-connection in $S[\omega , e]$ 
reproduces the result presented in \cite{Zinoviev:2005qp}. \\

On the branch \eqref{Einsteinlike} with 
$P_\mu{}^a = -\lambda^2/2 \, e_\mu{}^a\,$ that, in particular, 
contains AdS$_{D}\,$, we have that the fully 
nonlinear child action $S[\omega]=S[\omega, e(\omega)]$
assumes exactly the same value as the child action of the MacDowell--Mansouri
action \eqref{McDM}. Indeed, 
in the present case
the Weyl two-form becomes  
$C^{ab} = R^{ab} + \lambda^2\, e^a e^b\,$ which 
coincides with the AdS$_D$ curvature ${\cal R}^{ab}\,$.

\subsection{Adding the Pontryagin invariant}

\noindent 
In this subsection we adopt the Euclidean signature 
and add a topological invariant to the parent 
action \eqref{nonlinearCC} in $D=4\,$, 
the Hirzebruch signature $\tau(\mathscr{M}_4)$ of the 
manifold:
\begin{equation}
\tau(\mathscr{M}_4) = \tfrac{1}3 \int_{\mathscr{M}_4} p_1(\mathscr{M}_4)\ ,
\end{equation}
where $p_1(\mathscr{M}_4)$ the first Pontryagin 
class which, in this particular case, can be written as:
\begin{equation}
p_1(\mathscr{M}_4) = -\tfrac{1}{8 \pi^2} \epsilon_{abcd} \, R^{ab}\wedge *R^{cd} 
= -\tfrac{1}{8 \pi^2} \epsilon_{abcd} \, C^{ab}\wedge *C^{cd}\ ,
\end{equation}
where $*$ denotes the Hodge dual. Using the identity 
in Lorentzian Euclidean signature
\begin{equation}
\epsilon_{abcd}\, C^{ab}\wedge C^{cd} 
= \epsilon_{abcd}\, *C^{ab}\wedge *C^{cd}\ ,
\end{equation}
the action \eqref{nonlinearCC} in $D=4$ can be 
rewritten as $S[e,\omega] = -\tfrac{1}{4\lambda^2} 
\int_{\mathscr{M}_4} \epsilon_{abcd} \left( C^{ab}\wedge C^{cd} 
+ *C^{ab}\wedge *C^{cd} \right)\,$. \\
By adding to it a term proportional to the Hirzebruch invariant, 
we can recast the resulting action 
in a form that only depends on the (anti-) self dual 
part of the Weyl tensor. More precisely, one has
\begin{eqnarray}
S^\tau[e,\omega] & = & -\tfrac{1}{4\lambda^2} \int_{\mathscr{M}_4} \epsilon_{abcd} \left( C^{ab}\wedge C^{cd} + *C^{ab}\wedge *C^{cd} \right) 
\pm \tfrac{12 \pi^2}{\lambda^2} \tau(\mathscr{M}_4)
\nonumber \\ & = & -\tfrac{1}{4\lambda^2} \int_{\mathscr{M}_4} \epsilon_{abcd} \left( C^{ab}\wedge C^{cd} + *C^{ab}\wedge *C^{cd} \pm 
\left[ C^{ab}\wedge *C^{cd} + *C^{ab}\wedge C^{cd} \right] \right) 
\nonumber  \nonumber  \\ & = & -\tfrac{1}{4\lambda^2} \int_{\mathscr{M}_4} \epsilon_{abcd} \left( C^{ab} \pm  *C^{ab} \right) 
\wedge\left( C^{cd} \pm *C^{cd} \right) \nonumber \\ 
& = & -\tfrac{2}{\lambda^2} \int_{\mathscr{M}_4} e\, d^4x\,\sqrt{\text{det}\left(C^{ab} \pm *C^{ab}\right) }
\label{thetaterm}
\end{eqnarray}
where the last line is obtained by recalling the relation 
det$(A)=[$Pf$(A)]^2\,$ and after a small abuse of notation,
by extracting the volume form out of the wedge product 
of the 2-forms appearing on the second to last line. 
%
\\

A relation between the MacDowell--Mansouri action 
and Ashtekar's formulation with cosmological term 
was given in \cite{Nieto:1994rm}.
For discussions about MacDowell--Mansouri formulation of gravity 
in the context of S$\,$-duality, see for instance  
\cite{GarciaCompean:1998qh,GarciaCompean:1998wn,Julia:2005ze} and refs. therein.
It would be interesting to perform similar analyses starting from the 
alternative action \eqref{thetaterm}, where the difference is essentially 
that the Weyl two-form replaces the (A)dS$_D$ curvature two-form. 
It would also be interesting to reconsider the work \cite{Nieto:1996wn} 
starting from the action \eqref{thetaterm}.

\section{Conclusions}\label{sec:5}

When the spin-connection is dynamical and the cosmological constant is 
nonvanishing, 
the vielbein becomes an auxiliary field in the technical sense that it can be 
integrated 
out via its own algebraic equation of motion.
In the present paper, we analysed in details at quadratic level the corresponding 
pure spin-connection formulation of GR with a cosmological constant arising from 
the first-order MacDowell--Mansouri action.\\


We proved, using the unfolded technology, that the fluctuations of the 
dynamical spin-connection around (A)dS$_D$ propagate a massless 
spin-two particle.
We have also shown, by going to the traceless gauge, that the Hodge dual of the 
gauge field is a $GL(D)$-irreducible tensor field in the same representation as 
the dual massive graviton on flat spacetime of the same dimension.
Finally, the first-order quadratic action has been rewritten suggestively 
as the square of the linearised Weyl-like two-form. 
Interestingly, the nonlinear completion of this action obtained by inserting 
the full 
Weyl-like two-form is distinct from MacDowell--Mansouri's and Weyl 's actions,
though intimately related to both of them since they share the same values on 
Einstein spaces.\\

The nonlinear extension of the electric-magnetic duality of linearised gravity 
remains an important challenge  since, by analogy with its spin-one counterpart, 
such a duality should relate weak and strong coupling regimes.
In the presence of a cosmological constant, another type of duality is available 
between the conventional descriptions of gravity and exotic pure-connection 
descriptions.
One can see on dimensional ground that the loop expansion in any pure-connection 
formulation of gravity (Eddington--Schr{\"o}dinger's or pure spin-connection) is 
controlled by the ratio of the Planck length over the cosmological scale 
$|\Lambda|^{-\frac12}\,$.
As exhibited above at linearised level, these two types of dualities are deeply 
intertwined, if not even two faces of the same coin. 
Put together, these remarks suggest the existence of two dual descriptions 
of gravity with, respectively, 
small and large values of the cosmological constant.

\section*{Acknowledgements}

N.B. is supported by the ARC contract N$^{\rm o}$ AUWB-2010-10/15-UMONS-1.
He wants to thank the hospitality of the 
Laboratoire de Math\'ematiques et Physique Th\'eorique of Tours University, 
where he spent a visiting professor stay during which the present work was initiated.
T.B. is supported by a joint grant ``50/50'' R\'egion Centre / UMONS.

\providecommand{\href}[2]{#2}\begingroup\raggedright\endgroup


\end{document}